\documentstyle[preprint,aps,12pt,eqsecnum,epsf]{revtex}

\begin{document}

\thispagestyle{empty}

\title{
\begin{flushright}
{\normalsize DOE/ER/41014-40-N97 }\\
\hspace {3.7in} October 1997\\
\end{flushright}
Nucleon Charge Symmetry Breaking and Parity Violating Electron-Proton
Scattering}

\author{Gerald A. Miller}

\address{Department of Physics, \\
University of Washington, Box 351560, \\ 
Seattle, Washington 98195, USA
\vspace{15pt}
}

\maketitle

\begin{abstract}
The  consequences of the charge symmetry breaking effects   of the
mass difference between the up and down quarks and electromagnetic
effects for searches for strangeness form factors in parity violating
electron scattering from the proton are investigated. The formalism
necessary to  identify and compute the relevant  observables is
developed by separating  the Hamiltonian into charge symmetry
conserving and breaking terms. Using a set of SU(6) non-relativistic
quark models,  the effects of charge symmetry  breaking Hamiltonian
are considered for experimentally relevant  values of the momentum
transfer and found to be less than about 1\%. The charge symmetry
breaking corrections to the Bjorken sum rule are also studied and
shown  to vanish in first-order  perturbation theory.
\end{abstract}

\pacs{ none yet}
\newpage

\section{Introduction}

If one neglects the mass difference between the up and down quarks and
ignores electromagnetic effects, the QCD Lagrangian that governs hadronic
physics would be invariant under the interchange of
up and down quarks. This invariance is called charge symmetry, which is
more restrictive than isospin symmetry which involves invariance under
any rotation in isospin space.  Small, but
interesting, violations of charge symmetry have been discovered and are
described in the
reviews\cite{mns,mvo,others}. All charge symmetry breaking 
effects arise from the mass difference
between the up and down quarks and from electromagnetic effects.

The second European Muon Collaboration EMC
 effect\cite{spindata}, the discovery that valence quarks carry only a small
fraction of the nucleon spin,  and the resulting search
for strangeness in the nucleon  has brought some attention to understanding
the role of nucleonic charge symmetry breaking. 
If this symmetry holds,
measurements of a parity violating
electron
left-right asymmetry in electron-proton scattering can determine new form
factors whose origin
lies only in the strange and anti-strange quarks of the nucleon
\cite{km,mckdhb}.  However, the symmetry does not hold precisely and
it is of interest to estimate how small the effects can be. This is especially
true now that the first measurement of the proton's neutral weak magnetic
form factor finds a value of the strange magnetic form factor that is
consistent with zero\cite{sample}.

Another issue concerns the momentum transfer $Q^2$
dependence of any charge symmetry
breaking effects. In principle, the charge symmetry breaking terms,
 which act as a perturbing Hamiltonian, can cause the nucleon to mix with
states which would otherwise be orthogonal. Such components could cause
the form factor to have a $Q^2$ dependence which could emphasize the effects
of charge symmetry breaking. The purpose of this paper is to 
present arguments that such                                                    
a possibility can not occur.

It is worthwhile to discuss briefly how the assumption of
charge symmetry simplifies  the
analysis of parity violating electron scattering\cite{mckdhb}.
The difference in cross section for right and
left handed incident electrons arises from the interference of the photon and
Z-boson exchange terms. In particular, the  photon-electron coupling is vector
and the Z-electron coupling is axial, while the boson-proton coupling is
vector. The matrix element for 
Z-boson proton coupling, $
M_{fi}^\mu (Q^2)$ is given by\cite{musolf}
\begin{equation}
M_{fi}^\mu (Q^2) = 
\langle p,f \mid \bar{u} \gamma^\mu u - \bar{d} \gamma^\mu d 
\mid p,i \rangle
-\frac{1}{3} 
\langle p,f \mid \bar{s} \gamma^\mu s \mid p,i \rangle
 -4 \sin^2 \theta_W 
J_{p,fi}^\mu (Q^2). \label{one}
\end{equation}
Our notation is that the $\mid p,i\rangle$ 
denotes a proton in an initial state
with momentum and spin denoted by $i$. The terms $\bar{u} \gamma^\mu u$
and $\bar{d} \gamma^\mu d$
 are evaluated at the space-time origin. The electromagnetic matrix element 
of the proton is denoted as $J_{p,fi}^\mu (Q^2)$, and the nucleonic term
$N=p,n$  is
defined as 
\begin{equation}
J_{N,fi}^\mu (Q^2) \equiv 
\langle N,f \mid \frac{2}{3} \bar{u} \gamma^\mu u - 
\frac{1}{3} \bar{d} \gamma^\mu d - \frac{1}{3} \bar{s} 
\gamma^\mu s \mid N,i\rangle. \label{jmu}
\end{equation}
The second term of Eq.~(\ref{one}) is directly related to the
strangeness of the nucleon, and is the new feature of parity-violating electron
scattering. The third term of Eq.~(\ref{one}) is well measured, but
to extract the strange properties it is necessary to 
to determine the first term from independent experiments. 
We define this term as $X_{fi}^\mu (Q^2)$ with 
\begin{equation}
X_{fi}^\mu (Q^2) \equiv 
\langle p,f \mid \bar{u} \gamma^\mu u - 
\bar{d} \gamma^\mu d \mid p,i \rangle. \label{xdef}
\end{equation}
If charge symmetry holds, the (u,d) quarks in the proton are in the same wave
function as the (d,u) quarks in the neutron,  and the strange quark wave
functions of the neutron and proton are identical. In that case
\begin{equation}
X_{fi}^\mu (Q^2) = J_{p,fi}^\mu (Q^2) - J_{n,fi}^\mu (Q^2),
\end{equation}
and the right hand side is well measured. 
We aim to study the error involved in asserting that the equality 
holds exactly.

Here is an outline of this paper. The next section is  concerned
with displaying the  charge symmetry 
formalism which allows a definition of
the terms that cause the  charge symmetry breaking correction to
$X^\mu_{fi}(Q^2)$. This correction $\delta X^\mu_{fi}(Q^2)$ is
obtained as a specific matrix element involving the charge symmetry
breaking Hamiltonian. This formalism is general, but our application
involves the non-relativistic quark model. This model is well  enough
founded as to allow reasonable estimates of the charge symmetry
breaking effects and is simple enough so that some general
conclusions, that go beyond the specific calculations,  can be drawn.
Three different non-relativistic quark models are defined in Section
3. Computing the perturbative corrections to the form factors involves
summing over all of the unperturbed intermediate states. This sum can
be simplified by using an approximation in which the unperturbed
Hamiltonian can be treated as a number, an average excited state mass
$M^*$, so that the sum over states can be performed using closure.  The
mass $M^*$ can be chosen so that the first correction to the closure
approximation vanishes, with the result that $M^*$ depends on $Q^2$
and on the  perturbing Hamiltonian. This closure treatment is worked
out in Section 4. The charge symmetry breaking observables are
computed in Section 5. Section 6 discusses the charge symmetry
breaking correction to the Bjorken sum rule
\cite{bjsrl}. Section 7 is reserved for a summary and a
discussion of the implications of the calculations. In addition, a
comparison with  other theories of nucleonic charge symmetry
breaking\cite{pollock,bian,shakin} is presented.
\section{Charge Symmetry Formalism}

The isospin formalism is elaborated in several
reviews\cite{mns,mvo,others}. Here we apply it to the nucleon and to
the calculation of the quantity $
X_{fi}^\mu (Q^2)$. The starting point is to realize the approximate
invariance of the Lagrangian under the interchange of $u$ and $d$
quarks. This makes it worthwhile to define the charge symmetry
operator, which is an isospin  rotation by 180$^\circ$ about the
y-axis
(taking the z-axis to be associated with the charge). This is defined
by
\begin{equation}
P_{cs}^\dagger u P_{cs} = d,
\end{equation}
with 
\begin{equation}
P_{cs} = \exp{i\pi T_2},
\end{equation}
and
\begin{equation}
T_2={1\over 2} \bar{q} \tau_2 q,
\end{equation}
where $q$ is the light (u,d) quark field operator.

The Hamiltonian consists of a charge symmetry conserving term $H_0$
and a 
breaking term $H_1$  such that 
\begin{equation}
H = H_0 + H_1,
\end{equation}
with 
\begin{equation}
[H_0, P_{cs}] = 0,
\end{equation}
and 
\begin{equation}
[H, P_{cs}] = [H_1, P_{cs}]. \label{comm}
\end{equation}

The unperturbed states are denoted by a subscript 0 and defined  by
\begin{equation}
H_0 \mid p,i \rangle_0 = \sqrt{\bar{M}^2 + \vec{p}^2} \mid p,i \rangle_0
 = E_i 
\mid p,i \rangle_0,
\end{equation}
where $\bar{M}$ is the average of the neutron and proton masses. 
We work to first order in  $H_1$ such that the physical proton is
expressed in terms of the unperturbed states 
by 
\begin{equation} 
\mid p,i \rangle = \mid p,i \rangle_0 + \frac{1}{E_i-H_0} \Lambda_i H_1 \mid
p,i\rangle_0.
\end{equation}
The quantity $\Lambda_i$ is a projection operator on to states
orthogonal to the unperturbed ground state isospin -doublet:
\begin{equation}
\Lambda_i=I -\mid p,i\rangle\langle p,i\mid -\mid n,i\rangle\langle
n,i\mid.
\end{equation}

The measured electromagnetic matrix elements are then obtained using
first-order perturbation theory as 
\begin{equation}
J_{p,fi}^\mu (Q^2) = 
_0\langle  p,i \mid  (\frac{2}{3} \bar{u} \gamma^\mu u - 
\frac{1}{3} \bar{d} \gamma^\mu d)(1 + \frac{1}{E_i-H_0} 2 \Lambda_i H_1) 
\mid p,i \rangle_0, \label{jp}
\end{equation}
and
\begin{equation}
J_{n,fi}^\mu (Q^2) = 
_0\langle  n,f \mid (\frac{2}{3} \bar{u} \gamma^\mu u - 
\frac{1}{3} \bar{d} \gamma^\mu d)
(1 + \frac{1}{E_i - H_0} 2 \Lambda_i H_1) \mid n,i \rangle_0. \label{jn}
\end{equation}

We may relate the neutron and proton matrix elements 
by using charge symmetry which holds for unperturbed states:
\begin{equation}
\mid n,i \rangle_0 = P_{cs} \mid p,i \rangle_0,
\end{equation}
and  which also gives
\begin{equation}
P_{cs}^\dagger \bar{u} \gamma^\mu u P_{cs} = \bar{d} \gamma^\mu d.
\end{equation}
This along with Eq.~(\ref{comm}) allows one to obtain the relation
\begin{equation}
P_{cs}^\dagger H_1 P_{cs} = H_1 + \Delta H, \label{dh1}
\end{equation}
where
\begin{equation}
\Delta H \equiv P_{cs}^\dagger H P_{cs} - H.\label{dh2}
\end{equation}
This equation is useful in identifying the charge symmetry 
breaking parts of the Hamiltonian which are relevant here. In particular,
the isospin-vector operators are selected and doubled in taking the 
difference between the neutron and proton.
The evaluation of $\Delta H$ will proceed  by using the identity 
\begin{equation}
P_{cs}^\dagger\widehat{\tau_3 } P_{cs} = - \widehat{\tau_3},
\end{equation}
expressed in terms of field operators $\widehat 
\tau_3=\int d^3x\;\left(u(x)^\dagger u(x)-d(x)^\dagger d(x)\right)$. In 
first-quantized notation this is:
\begin{equation}
P_{cs}^\dagger \tau_3(i)  P_{cs} = - \tau_3(i).
\end{equation}
 
Using Eqs.~(\ref{jp}) and (\ref{jn}) and recalling the definition
(\ref{xdef}) of
the  relevant quantity $X_{f,i}^\mu (Q^2)$ which involves matrix elements of 
the physical proton state  leads to the desired result
\begin{equation}
X_{f,i}^\mu (Q^2) = J_{p,fi}^\mu (Q^2) - J_{n,fi}^\mu (Q^2) + \delta  
X_{f,i}^\mu (Q^2),
\end{equation}
where
\begin{equation}
\delta X_{f,i}^\mu (Q^2) \equiv 
_0\langle  p,f \mid (\frac{2}{3} \bar{d} \gamma^\mu 
d-\frac{1}{3} \bar{u} \gamma^\mu u) \frac{\Lambda_i}{E_i - H_0} 2 \Delta H 
\mid p,i \rangle_0 \label{main}
\end{equation}
Obtaining the equation for $\delta X_{f,i}^\mu (Q^2)$ is the main
result of the present formalism. This term contributes to the observed
parity violating signal in just the same way as the interesting
strangeness matrix element. It is therefore necessary to have some
understanding about its magnitude and its $Q^2 $ dependence.
The quantity $\delta X_{f,i}^\mu (Q^2)$ can be related to charge
symmetry breaking modifications of the form
factors $G_{E,M}$ or $F_{1,2}$. In particular, the $F_i$ so obtained
are the same as $1/2 \left(^{u-d}F_i^{p+n} -^{u+d}F_i^{p-n}\right)$ of 
 Dmitrasinovic and  Pollock\cite{pollock}.

\section{Non-Relativistic Quark Models}
The preceeding formalism is completely general. Here we adopt the view
that it is reasonable to 
use a set of non-relativistic quark models 
to understand the rough size of effects at low $Q^2$ and to make first
estimates of the $Q^2$ dependence. With these models, the necessary
evaluations are not difficult  and one gains insight into the physics
of charge symmetry breaking.

In non-relativistic quark models the spin and momentum 
 proton are not related so that we may specify our notation by the replacement
\begin{equation}
\mid p,i \rangle \to \mid p, \uparrow \rangle,
\end{equation}
for a spin up proton. The spin index will be treated implicitly so that 
$\mid p, \uparrow \rangle \to \mid p\rangle$.

The Hamiltonian is specified by a set of terms
\begin{equation}
H = K + V_{con} + V_{em} + V_g,
\end{equation}
including the kinetic energy operator $K$, the confining potential $V_{con}$
which
respects charge symmetry, and the residual electromagnetic $V_{em}$ and gluon
exchange $V_g$ interactions. 

We shall use Eq.~(\ref{dh2}) to identify the charge symmetry 
breaking Hamiltonian $\Delta H$ as a sum or contributions from the 
different terms of the Hamiltonian. Thus we shall obtain
\begin{equation}
\Delta H = \Delta  K + \Delta V_{em} + \Delta V_g.
\end{equation}
in which each term is obtained via the operation indicated in
Eq.~(\ref{dh2}) i.e. $\Delta K=P^\dagger_{cs}K P_{cs}-K$.

Specifically, the kinetic energy term is given 
by 
\begin{equation}
K = \sum_{i=1}^{3} (m_i + \frac{p_i^2}{2 m_i}), \label{ke}
\end{equation}
where $m_i$  depends on whether the i'th quark is an up or down quark. We use
the notation
\begin{equation}
m_i = \bar{m} + \frac{\Delta m}{2} \tau_3 (i),
\end{equation}
in which 
$\Delta m = m_u - m_d$.
Then 
\begin{equation}
K = K_0 + \Delta K,
\end{equation}
with 
\begin{equation}
K_0 = 3 \bar{m} + \sum_i \frac{p_i^2}{2 \bar{m}}
\end{equation}
and
\begin{equation}
\Delta K = \Delta m  \sum_i \tau_3 (i) + \frac{\Delta m}{\bar{m}} 
\sum_i \frac{p_i^2}{2 \bar{m}} \tau_3 (i).\label{dk}
\end{equation}
The first term of Eq.~(\ref{dk}) does not modify the unperturbed wave function
and is henceforth ignored.

The electromagnetic interaction contains charge symmetry breaking and
more general charge  dependent terms. This operator is given by 
\begin{equation}
V_{em} = \alpha\sum_{i<j} q_i\;q_j (\frac{1}{r_{ij}} - \frac{\pi}{2} \delta 
(\vec{r}_{ij}) [ \frac{2}{\bar{m}^2} + \frac{4}{3} \frac{\vec{\sigma}(i) 
\cdot \vec{\sigma} (j)}{\bar{m}^2} ]),
\end{equation}
where $
q_i = \frac{1}{6} + \frac{1}{2} \tau_3 (i)$
and $\vec{r}_{ij}  
\equiv \vec{r}_i - \vec{r}_j$. The charge asymmetric part of $V_{em}$ is 
given according to Eq.~(\ref{dh2}) as
\begin{equation}
\Delta V_{em} = - \frac{\alpha}{6} \sum_{i<j} ( \tau_3 (i) + \tau_3 (j) ) 
( \frac{1}{r_{ij}} - \frac{\pi}{\bar{m}^2} \delta ( \vec{r}_{ij} ) 
[ 1+ \frac{2}{3} \vec{\sigma} (i) \cdot \vec{\sigma} (j) ] ).\label{dvem}
\end{equation}

We take the gluon exchange operator to be 
\begin{equation}
V_g = -\alpha_s \sum_{i<j} \lambda_i \cdot \lambda_j [ 
\frac{\pi}{2} \delta  (\vec{r}_{ij}) (\frac{1}{m_i^2} + \frac{1}{m_j^2} + 
\frac{4}{3} \frac{\vec{\sigma} (i) \cdot \vec{\sigma} (j)}{m_i m_j}) ],
\end{equation}
where for three quark baryons:
$\lambda_i \cdot \lambda_j = - \frac{2}{3}$.
The long range $1/r_{ij}$ term respects charge symmetry and 
is not included here.
Such a term is included, in principle, as part of the flavor
independent confining interaction.
The charge symmetry breaking  piece of $V_g$ is given by 
\begin{equation}
\Delta V_g = \alpha_s \frac{2 \pi}{3} \frac{\Delta m}{\bar{m}^3} \sum_{i<j} 
(\tau_3 (i) + \tau_3 (j)) [ 1+ \frac{2}{3} \vec{\sigma} (i) \cdot 
\vec{\sigma} (j)] \delta (\vec{r}_{ij} ).\label{dvg}
\end{equation}

We note that the  short-range terms of the electromagnetic and gluon exchange
operators are rather similar, so that we may re-write the charge symmetry
breaking Hamiltonian as
\begin{equation}
\Delta H = \Delta K + \Delta V_L + \Delta V_S, \label{deltah}
\end{equation}
where
\begin{equation}
\Delta V_L = - \frac{\alpha}{6} \sum_{i<j} \left(\tau_3 (i) + \tau_3 (j)
\right)
\frac{1}{r_{ij}},
\end{equation}
and
\begin{equation}
\Delta V_s = (- \frac{\alpha}{6} + \frac{2}{3} \alpha_s 
\frac{\Delta m}{\bar{m}} )
\frac{\pi}{\bar{m}^2} \sum_{i<j} ( \tau_3 (i) + \tau_3 (j) ) 
\delta(\vec{r}_{ij} )
(1+ \frac{2}{3} \vec{\sigma} (i) \cdot \vec{\sigma} (j)).
\end{equation}
These expressions are used to simplify the evaluations performed in
the next section.

To proceed further we need to specify the confining potential and its
ground state wave function. We shall use oscillator confinement for
most of the calculations of this paper. Thus we write
\begin{equation}
 \mid p \rangle_0 = \mid \Psi\rangle
\frac{1}{\sqrt{2}} \left(\mid \phi_ S \rangle
\mid \chi_S\rangle + \mid \phi_A\rangle  \mid \chi_A
\rangle\right)\label{su6}
\end{equation}
where
\begin{equation}
\langle \vec{r}_i\mid \Psi\rangle=\Psi(\rho,\lambda)=
N e^{-(\rho^2 + \lambda^2) / 2 \beta}. \label{ho}
\end{equation}
Here $\vec{\rho}\equiv {1\over \sqrt{2}}(\vec{r}_1-\vec{r}_2)$ and 
$\vec{\lambda}\equiv {1\over \sqrt{ 6}}(\vec{r}_1 +\vec{r}_2 -2\vec{r}_3)$, 
and the dependence on the position of the center of mass
is not made explicit. 
Standard\cite{close} mixed symmetric spin ($\phi_S$) and isospin
($\chi_S$) wave functions are used. The mixed anti-symmetric ones are denoted
by the subscript $A$. If oscillator confinement is used, the 
full charge-asymmetric kinetic energy operator can be incorporated exactly 
into
the operator $H_0$. This is the procedure of Ref.~\cite{pollock}. We 
keep the first-order perturbative treatment here for two reasons. First, all
effects of first order in $m_d-m_u$ can be treated in the same way; and 
second we wish to go beyond the effects of oscillator confinement.
However, this difference in procedure does not lead to 
differences in the results of first order in $m_d-m_u$.

The above wave function can be used to compute the electric $G_E$
 and magnetic ($G_M$)
form factors. In the non-relativistic quark model these are given by 
the expressions:
\begin{equation}
G_E (Q^2) =   \langle p \mid \sum_i \frac{1}{2}(1 + \tau_3 (i)) 
e^{i \vec{q} \cdot \vec{r}_i} \mid p \rangle,\label{ge}
\end{equation}
and 
\begin{equation}
G_M (Q^2) = {\bar{M}}
\langle  p \mid \sum_i \frac{1}{2}(1 + \tau_3 (i)) 
{\sigma_3 (i)\over m_i} e^{i \vec{q} \cdot  \vec{r}_i} \mid p \rangle,
\label{gm}
\end{equation}
in which $Q^2=\vec{q}\cdot \vec{q}$.
These expressions need to be discussed because the 
the equations that relate 
$J^\mu_{N,f,i}$ of Eq.~(\ref{jmu}) to the form factors 
$G_E$ and $G_M$ depend
on the nucleon mass and one must therefore specify whether it the proton or
neutron mass or the average that enters. The discussion in 
\cite{pollock}  shows that in the Breit frame the 
quantity $G_M/M_N$ is proportional to the matrix element of the
quark magnetic moment operator $\sigma_3(i)/ m_i$. However the discussion
in Halzen \& Martin shows that the $1/M_N$ factor does not appear in the
definition of $G_M$, but that there is a factor of $M_N$ in the
definition of $G_E$. The difference arises because of different choices of
the normalization of nucleon spinors. One can not tell which is more
appropriate without doing a more complete treatment in which the
 nucleon-spinor representation is derived from the quark model. Since
 the spin and total momentum degrees of freedom are uncoupled, such 
 derivation is beyond the scope of the non-relativistic quark model.
 Thus 
we simply use the average mass $\bar M$ in Eq.~(\ref{gm}). This introduces 
a difference between our approach and that of Ref.~\cite{pollock}.
In principle, the differences are of order $(M_n-M_p)/\bar{M}\approx 1.3 \times
10^{-3}$ and ignorable\cite{exp}. We remind the reader that it is the 
quark mass difference, not the neutron proton mass difference, that sets
the scale of the charge symmetry breaking effects. The former quantity 
is larger
than the latter because it must compensate for 
the effects of the electromagnetic interaction and
the quark-mass dependence of the gluon exchange interaction 
which would cause the proton to be more massive than the neutron.
The values that $m_d-m_u$ might  take in different models are discussed next.

\subsection{Neutron-Proton Mass Difference and Model Parameters}

The parameters of the non-relativistic quark model  shall be
determined from the neutron proton mass difference and a consideration
of pionic effects. In first-order perturbation theory the mass
difference between the neutron and the proton can be expressed as a
matrix element of $\Delta H$: 
\begin{equation}
M_n - M_p = _0\langle  p \mid \Delta H \mid p \rangle_0.
\end{equation}
Evaluating the 
individual terms of Eq.~(\ref{deltah}) yields the following
results:
\begin{equation}
_0\langle  p \mid \Delta K \mid p \rangle_0 
= - \frac{(m_d - m_u)}{2 \beta \bar{m}^2},
\label{dkho}\end{equation}
\begin{equation}
_0\langle  p \mid \Delta V_{em} \mid p \rangle_0 = 
- \frac{\alpha}{3} \sqrt{\frac{2}{\pi \beta}} 
\left(1 - \frac{5}{12 \bar{m}^2 \beta} \right),
\end{equation}
and
\begin{equation}
_0\langle  p \mid \Delta V_g \mid p \rangle_0 = - 
\frac{\alpha_S (m_d - m_u)}{\bar{m}^3 \beta^{3/2}} 
\sqrt{\frac{2}{\pi}} \frac{5}{9}.
\end{equation}
Adding the individual terms leads to 
\begin{equation}
M_n - M_p = (m_d - m_u) \left[1- \frac{1}{2\beta \bar{m}^2} - \frac{\alpha_s}
{\bar{m}^3 \beta^{3/2}} \sqrt{\frac{2}{\pi}} \frac{5}{9} \right] 
- \frac{\alpha}{3} 
\sqrt{\frac{2}{\pi \beta}} \left(1 - \frac{5}{12 \bar{m}^2
\beta}\right).
\end{equation}

The parameters to be determined are $\beta,\alpha_s$, and $\bar{m}$.
We shall use $\bar m= 337 $ MeV as this leads to a proton magnetic
moment of 2.79 n.m.
The model used  does not include pionic effects because these are
essentially charge symmetric (as discussed below), but any
consideration of the parameters should take implicit account of the
pion cloud. We follow the ideas of the cloudy bag model\cite{cbm} in
which a perturbative treatment of pions as quantum fluctuations
converges for bag radii greater than about 0.6 fm. The importance of
pionic effects decreases as the bag radius  $R_B$ increases. The parameter
$\beta$ is essentially the mean square radius of the nucleon (which
corresponds to about 0.6 of $R_B^2$). We use the calculation of the
the $\Delta$-nucleon mass difference as a measure of pionic effects.
The gluonic contribution is given by:
\begin{equation}
(M_{\Delta} - M_N)_g = \frac{2}{3} \sqrt{\frac{2}{\pi\beta}} 
\frac{\alpha_s}{\bar{m}^2 \beta}.
\end{equation}
The physical value of this difference is taken here to be 300 MeV, but
pionic effects also contribute. So $
(M_{\Delta} - M_N)_g$ is a fraction $\gamma$ of 300 MeV. Larger values
of $\gamma$ correspond to 
 smaller pionic contributions and larger values of $\beta$. 
Three typical choices of parameters are shown in Table I.
We shall investigate the charge symmetry breaking using each of the three
models.

\section{Closure Approximation}

We are interested in computing  the charge symmetry breaking 
observables represented by 
Eq.~(\ref{main}). The different values of $\mu$ and the different helicities
specified by the quantum numbers $i,f$ can be used to specify the
contributions to the electric $E$ and magnetic $M$ terms. Separating
these terms  and using  
the non-relativistic wave function allows specifies Eq.~(\ref{main})
to 
\begin{equation}
\delta G_{E,M} (Q^2) = 
_0\langle  p \mid O_{E,M} (q) \frac{\Lambda}{\bar{M} - \bar{H}_0}
2\Delta H \mid p \rangle_0, \label{big}
\end{equation}
where
\begin{equation}
O_E (q) \equiv \sum_i \left( \frac{1}{6} - \frac{\tau_3 (i)}{2}\right) 
e^{i \vec{q} \cdot \vec{r}_i},
\end{equation}
and
\begin{equation}
O_M (q) \equiv \frac{\bar{M}}{\bar{m}} \sum_i \left( \frac{1}{6} - 
\frac{\tau_3 (i)}{2} \right)\sigma_3 (i) e^{i \vec{q} \cdot \vec{r}_i}.
\end{equation}
The operator $\bar{H}_0$ removes the center of mass 
kinetic energy operator from $H_0$:
\begin{equation}
\bar{H}_0 =H_0-{\left(\sum_i p_i\right)^2\over 2\sum_i m_i}.
\end{equation}
The expression (\ref{big}) depends only on internal coordinates
$\rho$ and  $\lambda$, so that the projection operator $\Lambda$ does
not depend on the initial and final nucleon momentum:
\begin{equation}
\Lambda=I-\mid p\rangle_0\;_0\langle p\mid 
-\mid n\rangle_0\;_0\langle n\mid.
\end{equation}

The evaluation of Eq.~(\ref{big}) depends on knowing the energies and 
wave functions of all of the eigenstates of $\bar{H}_0$. We shall
replace  $\bar{H}_0$ by a number $M^*_{E.M}(Q^2,\Delta H)$ 
which is expected to
depend on the momentum transfer, whether the electric or magnetic 
term is to be evaluated, and on the operator $\Delta H$.
 This quantity is determined from the condition that 
the  first correction to the simplification of the energy denominator 
by treating $\bar{H}_0$ as a number
vanishes. This determination is accomplished by adding and subtracting
 $M^*(Q^2,\Delta H))$ to $\bar{H}_0$:
\begin{equation}
\bar{H}_0 = M^*(Q^2,\Delta H) 
+ \left(\bar{H}_0 - M^*(Q^2,\Delta H)\right),
\end{equation}
and rewriting the energy denominator of Eq.~(\ref{main}) as
\begin{equation}
\frac{1}{\bar{M} - \bar{H}_0} \approx \frac{1}{\bar{M} - M^*(Q^2,\Delta H)} + 
\frac{1}{(\bar{M} - M^*(Q^2,\Delta H))^2} (\bar{H}_0 - M^*(Q^2,\Delta H)).
\label {approx}
\end{equation}
The requirement that the (unperturbed)
ground state expectation value of the second term of 
Eq.~(\ref{approx}) vanishes leads to the result:
\begin{equation}
M_{E,M}^* (Q^2 , \Delta H) - \bar{M} = \frac{1}{2} 
\frac{
_0\langle  p \mid [ [ O_{E,M} (q), \bar{H}_0 ], \Delta H ] \mid p \rangle_0}
{_0\langle  p \mid O_{E,M} (q) \Lambda \Delta H \mid p \rangle_0}.
\label{close}
\end{equation}
The use of the double commutator allows a straightforward evaluation
of the various average masses of the excited states. Observe that
the these masses depend on the operator $\Delta H$ and will be
different for the different contributions to $\Delta H$. It is convenient
to define 
\begin{equation}
\Delta E_{E,M} (Q^2, \Delta H) \equiv \bar{M} 
- M_{E,M}^* (Q^2, \Delta H),
\end{equation}
and also to use corresponding definitions for the individual
contributions to $\Delta H$.

The contributions to the electric terms can be obtained in a
straightforward manner. One simplification is that $_0\langle p\mid
O_E(q)\mid p\rangle_0=0$. Then
\begin{equation}
\delta G_E (Q^2) = 
\frac{_0\langle  p \mid O_E (q) 2 \Delta K \mid p \rangle_0}
{\Delta E_E (Q^2, \Delta K)}
+ \frac{_0\langle  p \mid O_E (q) 2 \Delta V_L \mid p \rangle_0}
{\Delta E_E (Q^2, \Delta V_L)}
+ \frac{_0\langle  p \mid O_E (q) 2 \Delta V_S \mid p \rangle_0}
{\Delta E_E (Q^2, \Delta V_S)}, \label{dge} 
\end{equation}
and
\begin{eqnarray}
\delta G_M (Q^2) = 
\frac{_0\langle  p \mid O_M (q)\Lambda 2 \Delta K \mid p \rangle_0}
{\Delta E_M (Q^2, \Delta K)}\nonumber\\
+ \frac{_0\langle  p \mid O_M (q)\Lambda 2 \Delta V_S \mid p \rangle_0}
{\Delta E_M (Q^2, \Delta V_S) } 
+ \frac{_0\langle  p \mid O_M (q)\Lambda 2 \Delta V_L \mid p \rangle_0 }
{\Delta E_M (Q^2, \Delta V_L)}
\label{dgm}
.\end{eqnarray}
%

The evaluation of the various terms $\Delta E_{E,M}(Q^2,\Delta H)$ is
a straightforward but tedious procedure, simplified by the feature
that only the $K_0$
part of $H_0$ contributed to the commutator
$[O_{E,M},\bar{H}_0]$\cite{lazy}.
Some of the relevant integrals are given in Table II.
Using the $\Delta K$ in the double commutator leads to the result
\begin{equation}
\Delta E_{E,M} (Q^2, \Delta K ) = - \frac{2}{\bar{m} \beta}=-2\hbar
\omega.\label{dek}
\end{equation}
That the above result must be obtained is an immediate consequence of the
oscillator confinement: the $p^2$ operator acting on the ground state
leads either to the ground state or to the $2 \hbar \omega$ excited
state. Here the procedure of evaluating the double commutator was
followed as a check on the algebraic procedure. 

The use of the long range $1/r_{ij}$ part of the  electromagnetic
operator leads to the following result for the related average
excitation energy:
\begin{equation}
\Delta E_{E,M } (Q^2, \Delta V_L ) = \frac{-1}{3 \bar{m}} 
\frac{\frac{5}{6} Q^2 e^{-Q^2 \beta /24} 
S_1 (Q^2\beta/2)+ \frac{e^{-Q^2 \beta /24}}{\beta} S_2 (Q^2\beta/2)}
{e^{-Q^2 \beta /24} S_1 (Q^2\beta/2) - e^{-Q^2 \beta /6}}, \label{long}
\end{equation}
where
\begin{equation}
S_1 (x) \equiv \sum_{n=0}^\infty ({-x})^n 
\frac{n!}{(2n+1)!},
\end{equation}
and
\begin{equation}
S_2 (x) = 4\;x\; \frac{dS_1}{dx} .
\end{equation}
Note that the average excitation energy turns out to be the same for
magnetic and electric probes. This is a consequence of the simple
wave functions employed and is related to the feature that
the electric and magnetic form factors have
the same $Q^2$ dependence.

The low momentum transfer limit, 
\begin{equation}
\lim_{Q^2 \rightarrow 0} \Delta E_{E,M} (Q^2, \Delta V_L) = 
- \frac{4}{\bar{m} \beta},
\end{equation}
shows that the $1/r_{ij}$ operator excites states of higher energy
than does the kinetic energy operator.

The using delta function contribution to $\Delta H$ leads to the 
following result:
\begin{equation}
 \Delta E_{E,M} (Q^2, \Delta V_S) =- \frac{5}{9} 
\frac{\frac{Q^2}{\bar{m}} e^{-Q^2 \beta /24}}
{e^{-Q^2 \beta /24} - e^{-Q^2 \beta /6}}, \label{short}
\end{equation}
and the low $Q^2$ limit is given by
\begin{equation}
\lim_{Q^2 \rightarrow 0} \Delta E_{E,M} (Q^2, \Delta V_S) = 
- \frac{40}{9 m \beta}.
\end{equation}
The latter expression shows that the delta function operator is the most
effective  (of the ones we consider)  at exciting the highest energy
states.

The terms $\Delta E_{E,M}$ depend only on the variable $Q^2\beta/2$.
If one multiplies $\Delta E_{E,M}$ by $\bar{m}\beta$ the result is a
function that is independent of the three models used here. This is
shown in Fig.~1. Note that $\bar{m}\beta$ decreases by a factor of
about two as one changes from model 1 to model 3. Thus model 1
corresponds to the smallest energy denominators. We shall display results
for $Q^2\beta/2\le 10$. Thus the maximum value of $Q^2$ is 
1.6, 2.2 and 3.1 GeV$^2/c^2$ for the models 1-3.
The planned parity violation experiments  are planned for 
values of $Q^2$ ranging from about 0.1 to  3 GeV$^2/c^2$ \cite{beck}.

\section{Charge Symmetry Breaking Observables}

We are now in a position to evaluate the effects of charge symmetry breaking 
for any value of $Q^2$. The charge symmetry breaking interactions $\Delta
V_{em}$ and $\Delta V_g$ include
two-body interactions that can be expected to lead to effects that fall off
slowly with increasing values of 
$Q^2$. We must compare such effects with the form factors $G_E$ and
$G_M$ computed in the limit in which charge symmetry holds. This is
because the gluon exchange
interaction $V_g$ includes a charge symmetric  term which will also lead
to slowly falling form factors. We will see that 
 this feature of the strong form factors
precludes a significant enhancement of charge symmetry breaking effects for
even the highest  values of $Q^2$ that we consider.                  
Thus the first task is to evaluate  
$G_{E,M}$ using the wave function $\mid p\rangle_0$.

\subsection{ $G_{E,M}(Q^2)$ With Charge Symmetry}
 
We shall evaluate $\mid p\rangle_0$ as arising from the harmonic 
confining potential including also 
the first-order effects of $V_g$. Starting with
perturbation theory is reasonable because the first-order 
contribution of $V_g$ to the nucleon mass is only -60 MeV for model 1 and
-25 MeV for model 3.
We shall see that for the range of $Q^2$ between 0 and 3 GeV$^2$/c$^2$
 relevant here
the influence of $V_g$ on the computed form factors can be 
reasonably large. This is  especially true for 
model 1 for which 
$\alpha_s =2.3$  as shown in Table I. We find
\begin{equation}
G_E (Q^2) = \exp{(-Q^2 \beta /6)} + \Delta G_E (Q^2), \label{geeq}
\end{equation}
with 
\begin{eqnarray}
\Delta G_E (Q^2) = - 4 \alpha_s 
 \frac{\pi}{3 \bar{m}^2} \frac{J_2 (Q^2)-J_2 (0)-2 (J_4 (Q^2)- J_4 (0))}
{\Delta E_e (Q^2, \Delta V_s)}  .\nonumber 
\end{eqnarray}
The integrals $J_i(Q^2)$ are tabulated in table II.

Similarly the magnetic form factor is obtained as:
\begin{equation}
G_M (Q^2) =\mu_p e^{-Q^2 \beta /6} + \Delta G_M (Q^2), \label{gmeq}
\end{equation}
where $\mu_p=2.79$ and 
\begin{equation}
\Delta G_M(Q^2) =  \alpha_s \mu_p  
 \frac{8\pi}{3 \bar{m}^2}
\frac{ \tilde{J}_4 (Q^2)}{\Delta E_m (Q^2, \Delta V_S)} ,
\end{equation}
with 
\begin{equation}
\tilde{J}_{3,4} (Q^2) \equiv J_{3,4} (Q^2) - J_{3,4} (0) e^{-Q^2 \beta /6}.
\end{equation}

The ratios $\Delta G_{E,M}/ G_{0E,M}$, where the form factors in the
absence of gluon exchange are given by $G_{0E}(Q^2)=\exp{(-Q^2\beta/ 6)}$
and $G_{0M}(Q^2)=\mu_P\exp{(-Q^2\beta/ 6)}$
are shown in Figs.~2 and 3. Both ratios
vanish at $Q^2=0$. Charge conservation mandates that this be so for the
electric form factor. However, the change in the magnetic term vanishes also
for  $Q^2=0$ because of the specific simplicities in the model unperturbed
wave function-
the spatially symmetric wave function multiplies the symmetric
 spin-isospin wave function. 
The correction $\Delta G_{E}$ is reasonably small, less than 20\% for all
of the values of $Q^2$ that we consider, but the magnetic correction,
 $\Delta G_{M}$ can be very large. If the absolute
magnitude ratio $\Delta G_{M}\over G_{,M}$ is larger than about 0.3, we can
expect that the perturbative treatment errs by more than about 10\%. Hence,
the largest values of ${Q^2\over 2\beta}$ for which the models can be
considered well defined are $\approx$ 5 and 7 for models 1 and 2. 
We will display the charge symmetry breaking form factors 
for values ${Q^2\over 2\beta}$ larger  than those limits
to provide information about the models, but the reader is cautioned against
taking those results seriously.

\subsection{Charge Symmetry Breaking}

We are now ready to evaluate the influence of charge symmetry on the measured
electric and magnetic form factors. We work to first order in perturbation 
theory (considering 
 the charge symmetry conserving one gluon exchange
interaction as a first order effect). The necessary equations  
(\ref{dge}) and (\ref{dgm}) are evaluated using the charge symmetry breaking
interactions of Eqs.~(\ref{dk}), (\ref{dvem}) and (\ref{dvg}). The average
excitation energies are given in Eqs.~(\ref{dek}),(\ref{long}) and
(\ref{short}). 

The evaluations are straightforward, so we simply express the results. 
We consider the influence of each charge symmetry breaking interaction 
$\Delta K$, $\Delta V_{em}$, and $\Delta V_g$ separately. 
Thus the contribution of $\Delta K$ to the electric form factor is given by
\begin{equation}
\delta G_E (Q^2, K) = - \frac{1}{9} \frac{\Delta m}{\Delta E_e (Q^2, K)}
\frac{Q^2}{\bar{m}^2} e^{-Q^2 \beta /6},
\end{equation}
while the magnetic form factor has a term
\begin{equation}
\delta G_M (Q^2, K) = - \frac{1}{27} \frac{\Delta m}{\Delta E_m (Q^2, K)}
Q^2 \beta e^{-Q^2 \beta /6}. 
\end{equation}
We see that the effects are order $\delta m/\Delta E\ll \delta m/\bar{m}$
times a small coefficient. Furthermore, the $Q^2$
dependence $ \delta G_{E,M} (Q^2, K)\sim Q^2 \beta e^{-Q^2 \beta /6}$ 
is different than that of the leading order dominant term $\sim \beta
e^{-Q^2 \beta /6}$ and this enhances the importance of charge symmetry
breaking at the higher values of $Q^2$ that we consider.

Including 
the effects of the electromagnetic interaction between quarks leads to the
following contributions to the form factors:
\begin{equation}
\delta G_E^{(em)} (Q^2) = - \frac{4 \alpha}{9} [\frac{(J_1 (Q^2) - J_3 (Q^2)}
{\Delta E_e (Q^2, \Delta V_L)} - \frac{\pi}{\bar{m}^2} \frac{5}{3}
\frac{(J_2 (Q^2) - J_4 (Q^2))}{\Delta E_e (Q^2, \Delta V_s)} ],
\end{equation}
and 
\begin{equation}
\delta G_M^{(em)} (Q^2) = \frac{8}{27} \alpha \mu_P 
\left[\frac{ 
 2 \tilde{J}_3 (Q^2)}{\Delta E_m (Q^2, \Delta V_L)} - 
\frac{\pi}{3 \bar{m}^2} 
\frac{
7 \tilde{J}_4 (Q^2)}{\Delta E_m (Q^2, \Delta V_s)}
\right].
\end{equation}
Here negligible effects are anticipated because of the small value of
$\alpha\approx 1/137$ and because of the large energy denominators.
These terms include the integrals $J_3$ and $J_4$ which fall much more slowly
than the leading order term, recall Table II.

Including
the effects of the gluon exchange 
interaction between quarks leads to the
following contributions to the form factors:
\begin{equation}
\delta G_E^{(g)} (Q^2) = 
\frac{\alpha_s}{\Delta E_e (Q^2, \Delta V_s)}
\frac{\Delta m}{\beta \bar{m}^3} 
\frac{20}{27} \sqrt{\frac{2}{\pi \beta}}
(e^{-Q^2 \beta /6} -e^{-Q^2 \beta /24} ),
\end{equation}
and 
\begin{equation}
\delta G_M^{(g)} (Q^2) = \frac{-4}{81} 
\frac{\alpha_s\mu_P}{\Delta E_m (Q^2, V_s)} 
\sqrt{\frac{2}{\pi}} \frac{\Delta m}{\bar{m}^3 \beta^{3/2}}
\frac{7}{2} (e^{-Q^2 \beta /24} - e^{-Q^2 \beta /6}) \label{dgmb}
\end{equation}
The explicit formulae show the appearance of the $e^{-Q^2 \beta /24}$ term
which, at higher values of $Q^2$
 is much 
bigger than the $e^{-Q^2 \beta /6}$ variation of the leading order
term. One might expect that this feature would allow the charge
symmetry breaking effects to stand out.
However, the leading order charge symmetric form factors also have 
a term, caused by gluon exchange, which also   varies as 
 $e^{-Q^2 \beta /24}$. 

The computed charge symmetry breaking electric form factors are shown in
Figs.~4 and 5 which display $\delta G_E/G_E$ as a function of $Q^2/2\beta $
 using  $ G_E$ of 
Eq.~(\ref{geeq}). Fig.~4 shows the three contributions to $\delta G_E/G_E$
arising, in model  1, from the individual charge symmetry breaking terms:
kinetic energy (K) electromagnetic interaction (em) and gluon
exchange (g). The electromagnetic term gives a negligible
contribution, but the other terms can give contributions that are as
large as 1\%.  The sum of the three contributions are shown in Fig.~5 
for each of the three models. The effects are largest for model 1
because of its large value of $\alpha_s$. It is possible that 
charge symmetry breaking could be as large as 2\%. If one wishes to
assert that only small values of $\alpha_s$ are allowed\cite{stan}, then the
maximum charge symmetry breaking would be about 1\%.

The computed charge symmetry breaking magnetic form factors are shown
in
Figs.~6 and 7. The ratio $\delta G_M/G_M$ is displayed as a 
function of $Q^2/2\beta $
 where $ G_M$ is given in
Eq.~(\ref{gmeq}). Fig.~6 shows the three contributions to $\delta
G_M/G_M$
arising, in model  1,  from the individual charge symmetry breaking
terms.  Once again, the electromagnetic term gives a negligible
contribution, but the  terms g and K can give contributions that are as
large as 1\%.  In this case the gluon exchange and kinetic energy
terms tend to cancel, with the sign difference arising from the different
combinations of spin matrix elements appearing in the magnetic terms. 
The net result shown in Fig.~7, 
for each of the three models, is that the largest effects 
are less than about 1\% for values of $Q^2/2\beta$ for which the models are
valid.

It is worthwhile to examine the low $Q^2$ effects  by determining the 
change in the mean square radii caused by the different terms. The  
unperturbed form factors each vary as $1-\beta Q^2/ 6$, and   $\beta$ is the
mean square radius. The charge 
symmetry breaking terms lead to behavior of the form 
 $1-(\beta+\delta\beta)Q^2/ 6$. We denote the various $\delta\beta$ 
according to whether related to the electric or magnetic terms and 
according to the origin of the effects.  The results are listed  in Table~III.
The electric terms are much bigger than the magnetic terms, for which the 
different terms tend to cancel. Thus only $\delta \beta_E$ is changed in a
non-negligible manner. For model 1, the sum of the individual
contributions gives for  model 1 a result ${\delta \beta_E\over\beta}=0.008$,
which corresponds to a 1.6\% change in the root mean square radius.

\subsection{Dependence on wave function}

The previous numerical results have been obtained using the harmonic
oscillator wave function. Are  the presently obtained  very small
values of the charge symmetry breaking effects a simple
consequence of this? Another way to ask this question is: Is it possible
to find a wave function for which the effects of charge symmetry
breaking are enhanced?

The purpose of this section is to address these questions through the
use of wave functions other than the harmonic oscillator. Such an
investigation is necessarily limited but will allow us to make arguments
that are more general.

We start by considering the simple wave function introduced by Henley \&
Miller (HM)\cite {hm90}. First the  SU(6) nature of the 3-quark wave
function spin-isospin wavefunction of Eq.~(\ref{su6})
is unchanged. Then
$\Psi(\rho,\lambda)$ is replaced by a function $\Psi(\rho^2+\lambda^2)$.
The generalization is to expand the
square of the wavefunction in terms of harmonic oscillator  wavefunctions
of the form given by Eq.~(\ref{ho}),
\begin{equation}
\Psi^2_{HM}(\rho,\lambda)=\int_0^\infty d\beta\;
g_{HM}(\beta)\;e^{-(\lambda^2+\rho^2)/ \beta}.
\end{equation}
Henley \& Miller chose the function $g(\beta)$ so that the resulting
 electric form
factor of Eq.~(\ref{ge}) is of the usual dipole type
$G_E(Q^2)={\Lambda^4\over (Q^2+\Lambda^2)^2}$. In this case
\begin{equation}
g_{HM}(\beta)(\pi\beta)^3={1\over 36}\Lambda^4\beta
\exp{(-\Lambda^2\beta/6)},
\end{equation}
and
\begin{equation}
\Psi^2_{HM}(R)={\sqrt{6}\Lambda^5
\over 108\pi^3R}K_1(\sqrt{2\over3}\Lambda R),
\end{equation}
with $R\equiv \sqrt{\rho^2+\lambda^2}$ and $K_1(x)$ is a Bessel function
of an imaginary argument.
This wavefunction was originally used along with a semi-relativistic
Hamiltonian in which the kinetic plus rest mass energy is given
by $\sqrt{p^2+m^2}$ and is not suited for calculations with the
non-relativistic 
operator of eq.~(\ref{ke}). This is because of the 
non-relativistic kinetic energy operator has an infinite expectation
value in the wave function $\Psi_{HM}$.
This very same wave function was also used in Ref.~(\cite{bil}).

We shall proceed here by using a different function $g(\beta)$, one
which  leads to a finite expectation values of the kinetic energy, but
which also leads to a power law falloff of the form factor. Using this
wavefunction will allow us to see if
the very small effects of charge symmetry are associated with the rapid
Gaussian fall off of form factors obtained 
from the oscillator model. In
particular we take
\begin{equation}
g(\beta)={\Lambda^8\over \pi^3\;6^5}
e^{-\Lambda^2\beta/6},
\end{equation}
which gives
\begin{equation}
\Psi^2(\rho,\lambda)=\int_0^\infty d\beta\;
g(\beta)\;e^{-(\lambda^2+\rho^2)/ \beta}, \label{intform}
\end{equation}
and therefore
\begin{equation}
\Psi^2(R)={2\sqrt{6}\Lambda^7\over
\pi^3\;6^5}R\;K_1(\sqrt{2\over3}\Lambda R).\label {cform}
\end{equation}
The integral form (\ref{intform}) is useful for evaluating matrix elements of
local operators such as $\exp{i\vec{q}\cdot\vec{r}_i}$ or $v(r_{ij})$. For such
operators the actions of taking the matrix element in a harmonic oscillator
wave function and integrating over $\beta$ commute, i.e.
one may integrate
the harmonic oscillator matrix element times $g(\beta)(\pi\beta)^3$ over 
$\beta$ to obtain the final answer. In particular, the evaluation of
Eq.(\ref{ge}) now is the integral 
of $\exp{(-Q^2\beta/6)}g(\beta)(\pi\beta)^3$ 
which leads to the result 
\begin{equation}
G_E(Q^2)=\left(\Lambda^2\over Q^2+\Lambda^2\right)^4. \label{slow}
\end{equation}

To see how this works in studying charge symmetry breaking 
effects, we evaluate the 
effects of the magnetic hyperfine  interaction. Recalling
Eq.~(\ref{dgmb}) we now have
\begin{equation}
\delta G_M^{(g)}(Q^2) = -\frac{4}{81}\frac{7}{2} 
\frac{\alpha_s\mu_P}{\Delta E_m (Q^2, V_s)}
\sqrt{\frac{2}{\pi}} \frac{\Delta m}{\bar{m}^3}
\int_0^\infty\;d\beta\; {\Lambda^8\over 6^5}
 \beta^{3/2}e^{-\Lambda^2\beta/6}
(e^{-Q^2 \beta /24} - e^{-Q^2 \beta /6}), \label{ddgm}
\end{equation}
which is evaluated as 
\begin{equation}
\delta G_M^{(g)}(Q^2) = -\frac{14}{81}
\frac{\alpha_s\mu_P}{\Delta E_m (Q^2, V_s)}
\sqrt{2}{3/4\over 6^{5/2}} \frac{\Delta m}{\bar{m}^3}
\left[{1\over (\Lambda^2+Q^2/4)^{5/2}}-{1\over(\Lambda^2+Q^2)^{5/2}}\right].
\label{dxgm}
\end{equation}
For large values of $Q^2$ this form factor falls roughly as 
$Q^{-5}$ which
 is slower than
the $Q^{-8}$ behavior of Eq.~(\ref{slow}). Thus it might seem that at large
enough $Q^2$  the 
charge symmetry breaking effects would dominate. This, of course,
 is not true.
The strong form factor of Eq.(\ref{gmeq}) has a term $\Delta G_M(Q^2)$
of the same momentum dependence as that of the charge symmetry breaking term
of Eq.~(\ref{dgm}). Thus the strong form factor would also have 
a $Q^{-5}$  behavior and would not be encumbered by the small factor 
$\frac{\Delta m}{\bar{m}}$.

There is a general  lesson that can be drawn from
 this exercise.  Small charge symmetry breaking effects
derived from a perturbative term in the strong Hamiltonian can not lead
to form factors of a different asymptotic form than that of the strong form
factors.

The only possibility to get new effects is from the charge 
symmetry breaking in the kinetic energy operator $\Delta K$.
One might think that the kinetic energy acting on $\mid \Psi\rangle$ might
generate a state vector with different behavior.
To assess the 
importance of $\Delta K$ the relevant expressions
of Sect.~IV must be re-evaluated using the wavefunction of Eq.~(\ref{cform}).
The calculations are tedious but straightforward, so  the results will
be presented after the model parameters are discussed.
The size parameter $\Lambda$ 
is chosen so that $G_E$ of Eq.~(\ref{slow}) is consistent with a
root mean square radius of 0.83 fm. This gives $\Lambda =$ 5.90 fm$^{-1}$.
One may compare the relevant size 
of our present
wave function with that of the harmonic oscillator in another manner
by computing the
contribution of the kinetic energy operator to the neutron-proton mass
difference 
\begin{equation}M_n-M_p=-{(m_d-m_u)\over 3\bar{m}^2} (4\pi)^2\int\rho^2
d\rho\int\lambda^2d\lambda{\lambda^2\over R^2}
\left({\partial\psi\over \partial  R}\right)^2, 
\end{equation}
which may be equated with the harmonic oscillator result of Eq.~(\ref{dkho}).
to obtain an equivalent harmonic oscillator parameter $\beta_{eq}$ 
in the latter such that $\sqrt{\beta_{eq}} =$ 0.77 fm. The present wave
function corresponds to a larger size than the oscillators used here. 
For purposes of estimation, we take $m_d-m_u$ = 5.2 MeV, which is the value of
model 1. A calculation of all  of the relevant charge symmetry breaking 
terms would probably lead to a value a bit larger than that because a value
of $alpha_s$ larger than the  2.3 of model 1 would be needed to reproduce the
$\Delta$-nucleon mass splitting. 

The results are shown in Figs.~8 and 9. The computed ratios of charge
symmetry breaking effects to charge symmetry conserving ones are displayed
as $\delta G_E/G_E$ or $\delta G_M/G_M$ as a function of $Q^2\beta/2$ where
$\sqrt{\beta}=0.77 $ fm. Here the electric and magnetic form factors  have the
functional form of Eq.~(\ref{slow}). Observe that the computed ratios are once
again very small.

\section{Bjorken Sum Rule}

The structure function $g_1(x,Q^2)$ can be measured  in lepton-nucleon
deep inelastic scattering DIS by using a polarized beam and a polarized
target\cite{spindata}. See the reviews\cite{spinrevs}. 
Here $x$ is the Bjorken variable.  The $Q^2$ dependence
of $g_1(x,Q^2)$ arises from perturbative QCD evolution effects and from
higher twist and target mass corrections. For the present purpose
 of evaluating
the influence of charge symmetry breaking  using non-relativistic
quark models it is sufficient to consider the Bjorken sum rule within
the framework of the naive parton model.

The naive parton model interpretation of the spin-dependent
DIS data is that the valence 
quarks 
contribute very little to the proton's spin. This startling finding 
motivated the
studies of parity  violation in electron-proton scattering
studies discussed here.
The parton model 
structure functions measure the probability for finding a quark with
momentum fraction $x$    
in the proton and which is polarized either in
the same $\uparrow$ or the opposite $\downarrow$
direction to the proton's polarization $\uparrow$, and 
the structure functions are described
by the four independent parton distributions
$ (q \pm {\overline q})^{\uparrow} (x);
 (q \pm {\overline q})^{\downarrow} (x).$ The function
$g_1(x)$ is given by:
\begin{equation}
g_1 (x) = {1 \over 2} \sum_q e_q^2 \Delta q(x),
\end{equation}
where
$
\Delta q(x) = (q^{\uparrow} + {\overline q}^{\uparrow})(x) -
			(q^{\downarrow} + {\overline q}^{\downarrow})(x)$
is the polarized quark distribution and $e_q$ denotes the quark charge.
                                    
In the naive parton model the integral
$$
\Delta q = \int_0^1 dx \ \Delta q(x)
$$
determines the fraction of the proton's
spin which is carried by quarks (and anti-quarks) of flavor $q$.
Thus, as reviewed recently\cite{spinrevs}, one obtains 
\begin{eqnarray}
\Gamma_{1,p}\equiv\int_0^1 dx \ g_1 (x) =
         {1 \over 18} \langle p, \uparrow\mid(4\Delta u + \Delta d+\Delta s)
\mid p,\uparrow\rangle.
\label{gammap}
\end{eqnarray}
We shall not explicitly write the proton spin $\uparrow$ in the
following development. Thus $ \mid p\rangle$ is to be understood as
denoting $\mid p,\uparrow\rangle $.
The operators $\Delta q$  are 
the axial current operators for the different quark flavors, $q$:
$ {\overline q} \gamma_{\mu} \gamma_5 q$.
The axial charge $g_A$ measured in beta decays is 
given by 
\begin{equation}
g_A = \langle  p \mid \Delta u - \Delta d \mid p \rangle.
\end{equation}

The Bjorken sum rule involves the neutron matrix element:
\begin{equation}
\Gamma_{1n} = \frac{1}{18} \langle  n \mid 4 \Delta u 
+ \Delta d + \Delta s \mid n \rangle.
\end{equation}
The use of the formalism of Sect.~II, and the fact that the 
strange quark field operator is not influenced by rotations in isospin
space, allows one to express this quantity as 
a proton  matrix element:
\begin{equation}
\Gamma_{1n} = \frac{1}{18}
\langle  p \mid 4 \Delta d + \Delta u+ \Delta s \mid p \rangle
+ \frac{1}{18} \: {_0\langle } p \mid (4 \Delta u + \Delta d )
\frac{\Lambda}{\bar{m} - H_0} 2 \Delta H \mid p \rangle_0.
\label{gamman}
\end{equation}
                      
Taking the difference between the equations (\ref{gammap}) and
(\ref{gamman}) leads to the result:
\begin{equation}
\Gamma_{1p} - \Gamma_{1n} = \frac{g_A}{6} + \Delta \Gamma,
\label{bjsr}
\end{equation}
where
\begin{equation}
\Delta \Gamma = - \frac{1}{18} \: {_0\langle } p \mid (4 \Delta u +
\Delta d)
\frac{\Lambda}{\bar{m} - H_0} 2 \Delta H \mid p \rangle_0.
\label{deltag}
\end{equation}                   
The first term of Eq.~(\ref{bjsr}) represents the Bjorken sum rule and
the second term is the naive parton model 
correction to it caused by charge symmetry
breaking.

We shall use the  non-relativistic quark models \cite {haha} to
estimate the size of $\Delta \Gamma$.
In this case the relevant operators are:
\begin{equation}
\Delta u = \sum_i  \frac{1 + \tau_3 (i)}{2}  \sigma_3 (i),
\end{equation} and 
\begin{equation}
\Delta d = \sum_i \frac{1 - \tau_3 (i)}{2} \sigma_3 (i)
\end{equation}
so that 
\begin{equation}
4 \Delta u + \Delta d = \frac{5}{2} \sum_i \sigma_3 (i) +
\frac{3}{2} \sum_i \sigma_3 (i) \tau_3 (i). \label{oper}
\end{equation}
The first term does not excite the nucleon and is irrelevant.
 The action of the second operator on the nucleon leads to
either a nucleon or  to a $\Delta$. This simplifies the calculation
enormously since only the $\Delta $ intermediate state needs to
 be 
included in the sum over intermediate states required to evaluate 
Eq.~(\ref{deltag}).

Then the expression for 
$\Delta \Gamma $ as obtained by using the relevant operator
(\ref{oper}) in the matrix element of Eq.~(\ref{deltag}) is simply
\begin{equation}
\Delta \Gamma =  \frac{1}{6} \:
{_0\langle } P \mid \sum_i \sigma_3 (i) \tau_3 (i)
\frac{\mid \Delta \rangle
 \langle  \Delta \mid}{M_\Delta - M_N} 2 \Delta H \mid p \rangle_0.
\end{equation}
Only the spin dependent pieces of the charge symmetry breaking
Hamiltonian $\Delta H$ can contribute to this matrix element. Thus
\begin{equation}
\Delta H \to - ( \frac{\alpha}{9} - \frac{4 \alpha_s}{9}
\frac{\Delta m}{\bar{m}} )
\frac{\pi}{\bar{m}^2} \sum_{i<j} ((\tau_3 (i) + \tau_3 (j) ) \delta
(\vec{r}_{ij}) \vec{\sigma}_i \cdot \vec{\sigma}_j.
\end{equation}
However, the spin dot products may each be replaced by unity because 
\begin{equation}
\vec{\sigma}_i \cdot \vec{\sigma}_j \mid \Delta \rangle =
\mid \Delta \rangle 
\end{equation}
for all pairs  $i,j$ of quarks. 
Furthermore the expectation value of the isospin operators vanishes.
Consider $i,j=1,2$  and note that 
\begin{equation}
\langle \Delta \mid \tau_3 (1) + \tau_3 (2) \mid N \rangle
= - ( \frac{4}{\sqrt{12}} )
\langle  \Delta \mid \phi_s u u d \rangle = 0.
\end{equation}
The spatial symmetry of the $\Delta $ and nucleon wave functions insures
that this vanishing occurs for all pairs $i,j$. 

The result of this calculation is that the influence of the charge
symmetry  breaking Hamiltonian on the Bjorken sum rule
vanishes. This exact 0 is due to the 
use of first-order perturbation theory within an SU(6) symmetric
wave function.

\section{Discussion}

Let's summarize. 
The charge symmetry breaking observables relevant for parity-violating
electron scattering and a general formalism for their evaluation are
obtained in Sect.~II. This formalism is just a simple way to keep track of the
effects of the charge symmetry conserving $H_0$ and violating $H_1$
Hamiltonians. The observables are evaluated using a set of 
three non-relativistic quark models,
each with harmonic oscillator confinement and obeying SU(6) symmetry,
that is defined in Sect.~III. The models are distinguished by their 
different size parameters, and are required to reproduce the  
$\Delta-$nucleon mass difference, or  a size-dependent fraction
thereof. The charge symmetric breaking effects included are the 
effects of the mass difference between the up and down quarks in the
kinetic energy operator and one-gluon exchange interaction, and 
 the electromagnetic interaction. One obtains a reasonable range of 
values of $m_d-m_u$ needed to reproduce the observed value of the
neutron-proton mass difference.

The charge symmetry breaking effects are  small and therefore
well-treated using first-order perturbation theory. One must sum 
over an infinite set of intermediate states to carry out the necessary
calculations. This summation is aided by the approximation of treating 
$H_0$ appearing in the energy denominator 
as a constant. The relevant  constant $\Delta E$ 
is chosen  so that the first correction
to the approximation vanishes; see Eq.~(\ref{close}).
This means that $\Delta E$ depends on the momentum transfer and the 
operator that excites the proton. This use of a constant allows one to
use closure to perform the sum over intermediate states. This
procedure is the subject of Sect.~IV.

The evaluations are presented in Sect. ~V. First the 
electric and magnetic form factors of the eigenstates of $H_0$ are obtained. 
The strong one-gluon  exchange operator gives a high momentum tail which
dominates the Gaussian term obtained from the oscillator wave function.
Then the influence of the 
three  charge symmetry breaking terms in the Hamiltonian
are evaluated for the three different models. The effects due to the u-d quark
mass difference are larger than that of the electromagnetic interaction,
but are themselves very small.
The largest of the effects we find are of the order of 1\% for the change in 
$G_E$ caused by charge symmetry breaking effects. Some larger values are 
shown in the figures, but these are for values of the momentum        
transfer which are outside of the regime of applicability of the
models we use. These small values arise because of the small sizes of
the basic effects: the ratio of the quark mass difference to
constituent quark mass is about 1/70 and, the tail caused by 
the  strong one gluon exchange potential makes it impossible to find
a region of
momentum transfer for which these effects can stand out.
This result is not a consequence of the use of oscillator
confinement. A different wave function, in which the square of the
wave functions is an integral of harmonic oscillator wave functions,
is also used, and very small  effects of charge symmetry breaking are
obtained.

The charge symmetry breaking correction to the Bjorken sum rule is
examined in Sect.~VI. Here the use of SU(6) symmetric wave functions
is shown to lead to a vanishing correction in first-order perturbation
theory.  Charge symmetry breaking therefore has no impact on current
studies of the validity of the Bjorken sum rule.

Next consider other computations of  the effects of charge
symmetry  breaking on the nucleon. The present work is most similar to
that of Ref.\cite{pollock}, and our results are consistent with
those,
except for one detail (see Sect.~III)
that depends on issues beyond the scope of the
non-relativistic quark model.
That  previous calculation is  extended here by including 
the effects of the strong and electromagnetic
hyperfine interaction, by studying the momentum transfer dependence and by
using a non-oscillator proton wave function.
One
difference 
is that in
Ref.\cite{pollock} the sum over intermediate excited states is
saturated by the $\Delta (1550)$. We use closure to carry out the sum.
The charge symmetry breaking operators are
isovector which act on T=1/2 states so that the intermediate states
can have either T=1/2 or T=3/2. The closure approximation used here allows the 
T=1/2 states to be included.

Let's also discuss  the work of Ma\cite{bian} who 
who presents his results in the form $-\delta
G^s_M\approx 0.006\to 0.088 $ n.m.. This is small compared to the current
experimental error of about 0.2 n.m., but relatively important compared to
the rather small strange magnetic moment
$G_M^s$=-0.066 n.m. from 
the baryon-meson fluctuation model of Ref.~\cite{sbma}.
The abstract states that 
the neutron proton mass difference of leads to an excess
of n = $\pi^-$p over p = $\pi^+ n$ fluctuations, but 
two different
effects actually lead to the results. 

The first is claimed to
arise from the light cone treatment of the non-interacting propagator.
If the light cone treatment is used, a term $P^+P^- -{\cal M}^2$ with
${\cal M}^2=\sum_{i=1}^{2} {k_{\perp i}^2+m_i^2\over x_i}$ replaces
our non-relativistic inverse propagator $\bar M -\bar {H}_0$. The value of
$P^+P^-= M_N^2$, the square of the nucleon mass. In Ma's treatment
this takes on the two values of $m_p^2$ or $m_n^2$.  This leads to a
numerical result $r^\pi_{p/n}=P(p=\pi^+n)/P(n=\pi^-p)=0.986$, which
corresponds to an excess of 0.2\% of n = $\pi^-$p fluctuations if $
P(p=\pi^+n)\approx P(n=\pi^-p)\approx0.15 $.
However, one should perform a light-cone
perturbation theory treatment of the charge symmetry breaking, which
involves treating the Hamiltonian operator $P^+P^-$ as a sum of charge
symmetry breaking and charge symmetry conserving terms. In this case
the relevant eigenvalue,  analogous to $\bar M$ used here in Sect.~II,
must be ${1\over 2} (m_p^2+m_n^2)$. Using this value and changes the
above result of 0.986 to 0.992; the effect is reduced by a factor of
2.

Actually,  the biggest effect used by Ma is caused by the assumption
that the radius
$R\sim 1 $ fm  of the $n\pi^+$ component of the proton is 2.5\% smaller
than that of the $p\pi^-$ component of the neutron. This according to
\cite{bian} could be caused by Coulomb effects. However the effects of the
Coulomb
potential and electromagnetic interactions in loop graphs is 
of order $\alpha/\pi\ll .025$, so that Ma's effect, while physically
reasonable,
is estimated to have too large a value. 
A reasonable estimate of the effect is could be 0.03 n.m\cite{priv}.
This
effect is too small to be relevant to experiments.

Finally, consider  the work of Celenza and Shakin\cite{shakin} who
computed the deep inelastic structure functions of the nucleon using a
quark model which preserves translational invariance.   Effects of
charge symmetry breaking enter in their calculation of  the ratio of
$F_2^n(x)/F_2^p(x)$. They can reproduce the experimental values of
this ratio by  allowing the neutron confinement radius to be about 10
percent larger than the corresponding proton radius. Such an effect is
well motivated, but the 10\% value is much larger than the
$< 1$\% effects found here.
                                                      
The net result is that the effects of charge symmetry
breaking on the nucleon wave function can be expected to be very
small. Only a limited number of models are discussed in the
present paper, but it seems very difficult to construct a reasonable
model of the nucleon which incorporates large charge symmetry
breaking effects. That any charge symmetry violating effect in the
Hamiltonian $H_1$ has its analog in the symmetry preserving
Hamiltonian $H_0$ is  a model independent statement. Thus the 
present 
conclusion about the lack of import of charge symmetry breaking
effects seems to be true in  independent of the particular model used.

This work was
stimulated by a talk given by D.H. Beck at the national INT.
I thank A. Bulgac for making some very useful comments. This work is
partially supported  by the USDOE.

\begin{figure}
 \caption{Energy denominators vs $Q^2\beta/2$. The term $m$ of  the figure 
is $\bar m$ of the text. Solid-energy denominator for the long range operator
Eq.~(\ref{long}). Dashed-energy denominator for the short range
operator Eq.~(\ref{short}).} 
\end{figure}
\begin{figure}
\caption{Changes in electric form factors due to $V_g$. The ratio of the
second $\Delta G_M$ to first $\mu_p e^{-Q^2\beta/2}\equiv G_{0M}$ terms of
Eq.~(\ref{gmeq}). The numbers refer to the models 1-3.}
\end{figure}
\begin{figure}
\caption{Changes in magnetic form factors due to $V_g$. The ratio of the
second $\Delta G_E$ to first $ e^{-Q^2\beta/2}\equiv G_{0E}$ terms of
Eq.~(\ref{geeq}). The numbers refer to the models 1-3.}
\end{figure}
\begin{figure}
\caption{Charge symmetry breaking electric form factor. The different
contributions are shown for model 1}
\end{figure}
\begin{figure}
\caption{Charge symmetry breaking electric form factors for each of the three
models.}
\end{figure}
\begin{figure}
\caption{Charge symmetry breaking magnetic form factor. The different
contributions are shown for model 1}
\end{figure}
\begin{figure}
\caption{Charge symmetry breaking magnetic form factors for each of the three
models.}
\end{figure}
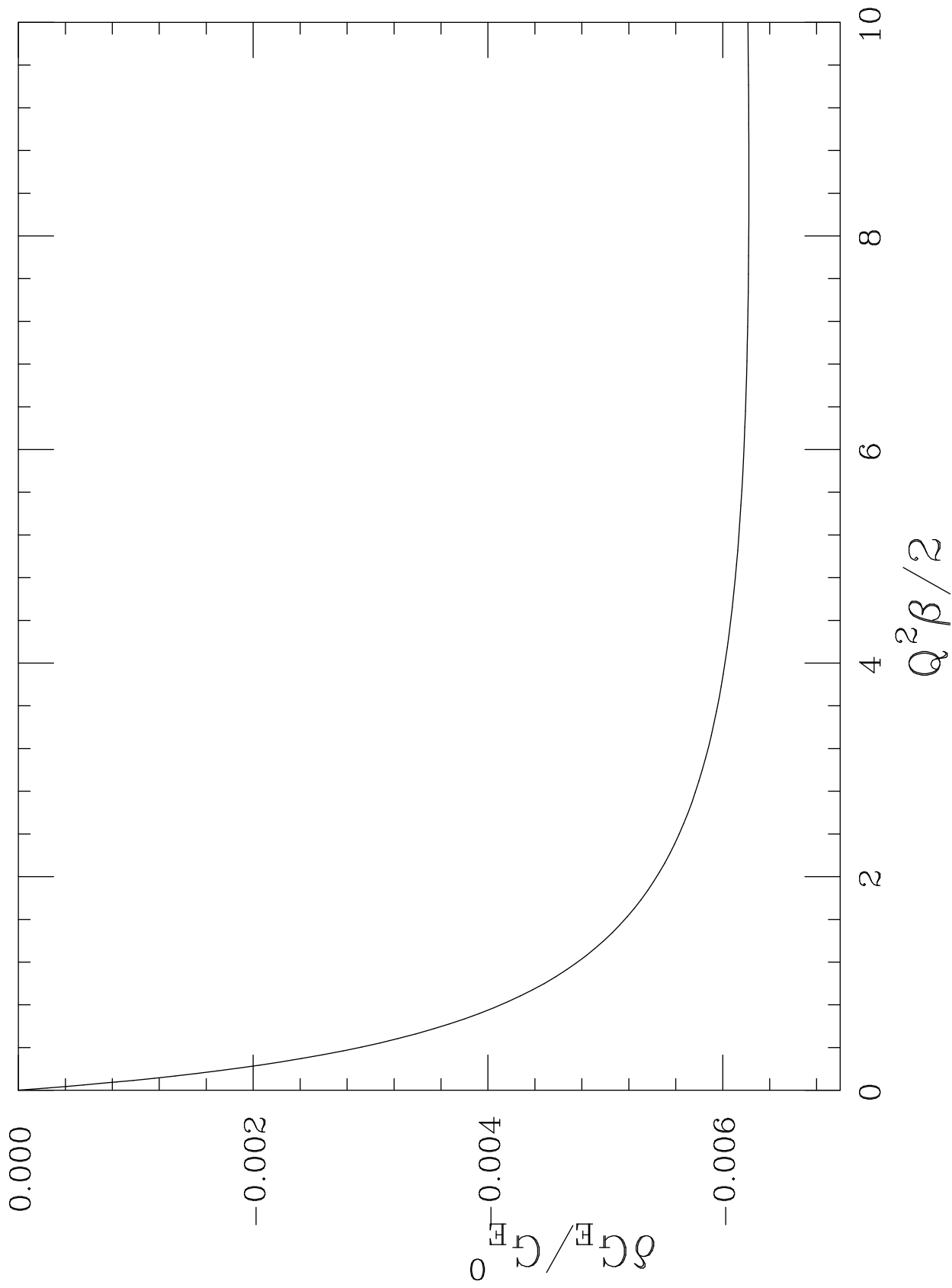
\begin{figure}
\caption{Use of the wave function of Eq. (5.16). Change in electric
form factor.}
\end{figure}
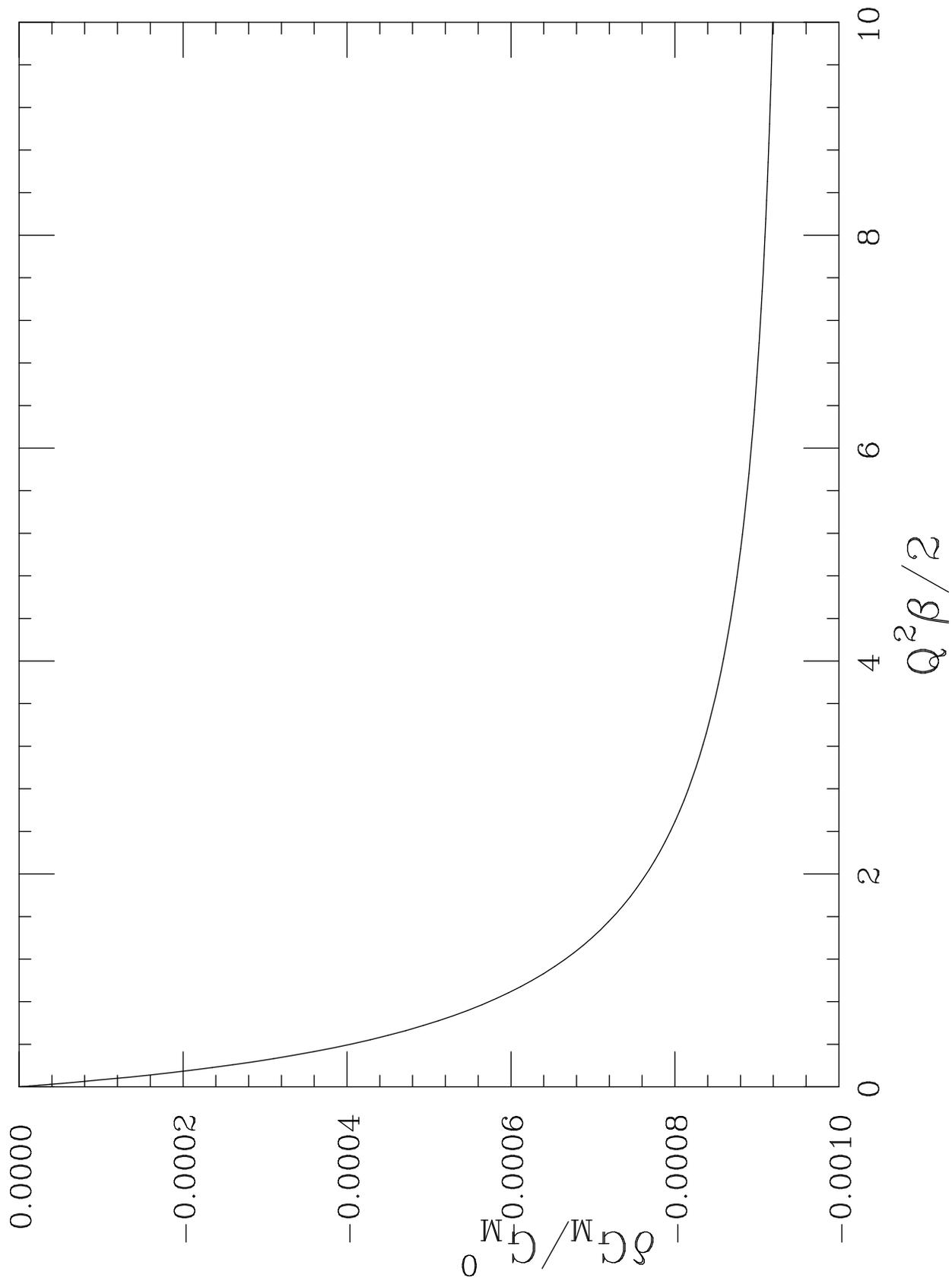
\begin{figure}
\caption{Use of the wave function of Eq. (5.16). Change in magnetic
form factor.}
\end{figure}

\begin{table}
\caption{Parameters of the Non-Relativistic Quark Models}
\begin{tabular}{l|rrr}
Model   & 1 & 2 & 3\\ \tableline
 $ \sqrt{\beta}$ (fm)  & 0.7 & 0.6  & 0.5 \\
$\sqrt{\beta}\bar{m}$  & 1.20 & 1.02 & 0.85\\
  $\alpha_s$  & 2.3 & 1.20 & 0.35 \\
  $m_d - m_u $ (MeV)  & 5.2 & 3.8 & 2.3 \\
  $\gamma$   &0.80 & 0.67 & 0.33 \\
\end{tabular}
\end{table}

\begin{table}
\caption{Relevant Integrals}
\begin{tabular}{llr}
  $J_i (Q^2) \equiv \int d^3 \rho\; d^3 \lambda \mid\psi(\rho,
  \lambda) \mid^2
   e^{i\vec{q} \cdot \vec{r}_3} 
  O_i $ \\ \tableline
  $O_1=\frac{1}{r_{12}}$ & $J_1 (Q^2) = \sqrt{\frac{2}{\pi \beta}} 
  e^{-Q^2 \beta /6}$ \\
  $O_2=\delta (\vec{r}_{12})$ &$ J_2 (Q^2)  =\frac{1}{4} 
  \sqrt{\frac{2}{\pi^3 \beta^3}} e^{-Q^2 \beta /6}$ \\
  $O_3= \frac{1}{r_{13}} $&$ J_3 (Q^2) = \sqrt{\frac{2}{\pi\beta}} 
  e^{-Q^2 \beta / 24}S_1(Q^2\beta/2)$
\\
 $ O_4 = \delta (\vec{r}_{13})$ & $J_4 (Q^2) = \frac{1}{4} 
  \sqrt{\frac{2}{\pi^3 \beta^3}} e^{-Q^2 \beta / 24}$
\end{tabular}

\end{table}
\begin{table}
\caption{Charge symmetry breaking changes in mean square radii, $\Delta
m=m_u-m_d$}
\begin{tabular}{l|rrr}
Cause
& $\Delta K$ & $\Delta V_{em}$ & $\delta V_g$\\ \tableline
${\delta \beta_E\over \beta}$ & 
 $
{-1\over 3}{\Delta m\over \bar{m}} $ &
${\alpha\over 
36}\bar{m}\sqrt{2\beta\over\pi}\left(1-{9\over 
8}{1\over\bar{m}^2\beta}\right)$
&
${-\alpha_s\over 8}{\Delta m\over \bar{m}} 
{1\over\bar{m}\sqrt{\beta}}\sqrt{2\over\pi}$
\\ \tableline
model 1 &0.0051  &-1.14$\times10^{-5}$ &0.0029 \\
model 2 &0.0038 &1.67 $\times10^{-5}$  & 0.0014\\
model 3 &0.0022 &4.38 $\times10^{-5}$ &0.0003 \\ \tableline
Cause  & $\Delta K$ & $\Delta V_{em}$ & $\delta V_g$\\ \tableline
${\delta \beta_M\over \beta}$
 & ${-1\over 9}{\Delta m\over \bar{m}}\bar{m}^2\beta $
&${-5\alpha\over 27}\bar{m}\sqrt{2\beta\over\pi}$ &
${-7\alpha_s\over 240}{\Delta m\over \bar{m}} 
{1\over\bar{m}\sqrt{\beta}}\sqrt{2\over\pi}$ \\ \tableline
model 1 &0.0024 &-0.0013 &-0.00069 \\
model 2 &0.0013 &-0.0011 &-0.00031 \\
model 3 &0.0005 &-0.0004 & -0.00007\\ \tableline
\end{tabular}
\end{table}


\begin{thebibliography}{99}
\bibitem{exp} The physical nucleon masses are treated as either
$2m_u+m_d$ or $m_u+2m_d$ in Ref.\cite{pollock}.
\bibitem{lazy}
We do not include the effect of the charge symmetric part of $V_g$ on
the 
evaluation of the quantities $\Delta E_{E,M}$. This enters in the
computation of $\mid p\rangle_0$. The inclusion of this
effect would decrease these quantities (increase the energy
denominator)
and decrease the effects of charge symmetry breaking). The results of 
Sect. V will show that such are very small. A further
reduction caused by an improved evaluation of $\Delta E_{E,M}$ would
have no consequence.
\bibitem{haha} The use of the non-relativistic quark model in which the 
proton is treated as three quarks to evaluate the very
small corrections to  a huge effect which seems to rule out the three-quark
idea may seem odd. However, this evaluation allows us to make 
a simple estimate and also to rule out the slim possibility that the
interesting results for $g_1(x)$ are caused by an unusually 
large charge
symmetry  breaking effect.
\end{thebibliography}
\end{document}